\documentclass[prb,twocolumn,showpacs,preprintnumbers,amsmath,amssymb,superscriptaddress]{revtex4}
\usepackage{graphicx}
\usepackage{bm}
\usepackage[usenames, dvipsnames]{color}

\begin{document}


\title{Skyrmion quantum spin Hall effect}

\author{Tianqi Chen}
\affiliation{Science and Math Cluster, Singapore University of Technology and Design, 8 Somapah Road,  487372 Singapore}
\affiliation{New York University Shanghai, 1555 Century Ave, Pudong New District, Shanghai 200122, China}

\author{Tim Byrnes}
\affiliation{State Key Laboratory of Precision Spectroscopy, School of Physical and Material Sciences, East China Normal University, Shanghai 200062, China}
\affiliation{New York University Shanghai, 1555 Century Ave, Pudong New District, Shanghai 200122, China}
\affiliation{NYU-ECNU Institute of Physics at NYU Shanghai, 3663 Zhongshan Road North, Shanghai 200062, China}
\affiliation{National Institute of Informatics, 2-1-2 Hitotsubashi, Chiyoda-ku, Tokyo 101-8430, Japan}
\affiliation{Department of Physics, New York University, New York, NY 10003, USA}

\date{\today}

\begin{abstract}
The quantum spin Hall effect is conventionally thought to require a strong spin-orbit coupling, producing an effective spin-dependent magnetic field.  However, spin currents can also be present without transport of spins, for example, in spin-waves or skyrmions. In this paper, we show that topological skyrmionic spin textures can be used to realize a quantum spin Hall effect.  From basic arguments relating to the single-valuedness of the wave function, we deduce that loop integrals of the derivative of the Hamiltonian must have a spectrum that is integer multiples of $ 2 \pi $.  By relating this to the spin current, we form a new quantity called the quantized spin current which obeys a precise quantization rule.  This allows us to derive a quantum spin Hall effect, which we illustrate with an example of a spin-1 Bose-Einstein condensate.  
\end{abstract}


            
\maketitle

\section{\label{sec:introduction}Introduction}
The quantum Hall effect (QHE) in its various forms -- the integer, anomalous, fractional, spin -- is now a cornerstone of physics and not only is of fundamental interest in solid-state physics \cite{PhysRevLett.95.226801,laughlin,IQHEexp,tsui,klinovaja2015integer,Chang167}, but also is actively researched in other fields such as atomic, optical, and high-energy physics 
\cite{beeler2013spin,hafezi,viefers,DTong,string}. 
In particular, the quantum spin Hall effect (QSHE) has attracted a huge amount of interest recently. One of the attractive aspects of the QSHE in comparison to the QHE is that no magnetic field needs to be applied to obtain the characteristic spin current flow without dissipation -- a potentially important effect for applications such as spintronics.  The theoretical prediction of its existence in HgTe quantum wells \cite{bernevig2006quantum} and fast experimental verification \cite{konig2007quantum} has spurred on intense research into related phenomena into topological insulators, topological superconductors, and other topological quantum states of matter \cite{TI,TI_TSC,fractional,konig2008quantum,qi2010quantum,diaz2019}. This has also motivated many experimental studies of spin 
Hall effects in various other physical systems \cite{kato2004observation,wunderlich2005experimental,hosten2008observation,RevModPhys.87.1213}.  To date, HgTe remains the only experimentally observed system where the QSHE has been observed. Part of the difficulty is that a large spin-orbit coupling is required to produce the QSHE edge states separating the spin channels, which is only strong enough in heavy elements such as HgTe.  

There is however another route to realize the QSHE. The QSHE via spin-orbit coupling relies upon a physical transport of carriers with spin in edge channels.   There is another type of spin current -- such as in spin waves or skyrmions --  where the spins are fixed in space, but their orientation varies spatially \cite{kajiwara2010transmission} [see Fig. \ref{fig1}(a)(b)]. Since these are equivalent ways of realizing a spin current, it should therefore be possible to realize the QSHE equally in this way, rather than in the conventional spin-orbit coupling approach. 
The concept of spin current has drawn more and more attention in recent years, where the manipulation of the spin degree of freedom has been explored in numerous works \cite{rashba2003spin,sun2005definition,shi2006proper,an2012universal,balachandran2018perfect}. Many efforts in probing spin currents have also been made in the field of spintronics \cite{romming2015field,zheng2017direct} and solid state physics \cite{maccariello2018electrical,li2016direct}. The definition of spin current operator has been proposed for systems without spin-rotational symmetry \cite{RuckriegelKopietz}. 

Recently, the skyrmion Hall effect was realized where a topological Magnus force was observed on the skyrmions \cite{jiang2017direct,litzius2017skyrmion,chen2017spin} as well as topological and spin Hall effects in disordered skyrmion textures \cite{Manchon2017}.  The control and detection of skyrmions in a Hall configuration is an essential first step towards realizing the quantum Hall effect. Furthermore, edge states from the QSHE are known to be closely related to the observation of spin current \cite{takahashi2008spin}. In this way, how one would be able to extend these results towards the quantum spin Hall effect with skyrmions is an important question. 

In this paper, we examine the realization of the QSHE using skyrmionic spin textures, instead of spin-orbit coupling. Our approach is to examine the general properties of spin textures, and topological observables that can be assigned to them.  The key idea in all QHE variants is to compare the longitudinal current to a transverse asymmetry, typically the potential difference.  Done correctly, this should depend on a topological invariant.  In our case, this has to do with the topology of the spin texture -- a related quantity to skyrmion number. 
A related approach was previously investigated in realizing the QHE in Bose-Einstein condensates (BECs) \cite{PhysRevA.92.023629}.  The key idea in the approach was to take advantage of the topological properties of vortices in such systems to observe quantized plateaus in the spatial and momentum distributions of the atoms.  Here we extend these ideas to arbitrary spin textures. Another related question is the generalization of the QSHE to higher symmetries.  Higher symmetries of the QSHE, such as the SU(3) QHE in SnTe \cite{PhysRevLett.116.026803,PhysRevLett.115.166805} also have attracted much interest in recent years. A relevant question is the realization of the QSHE in systems beyond the simple spin-up and spin-down cases as observed in HgTe. The separation of the spin-up and spin-down channels is only one manifestation of spin current that can be realized, and even for the spin-$1/2$ case more complex variations of the spin current should be feasible.  We show that in our formulation of the QSHE the scheme naturally generalizes to $\text{SU} (N)$, realizing more complex realizations of the QSHE. 



\section{\label{sec:spintexture} Realization of spin texture}
Our basic argument is as follows.  Consider a continuous spin texture in two dimensional space, with coordinates $ \bm{x} = (x,y)$. At each position $ \bm{x} $, an $N$-dimensional spin exists in a state $ | \psi(\bm{x}) \rangle $. A spin texture is then defined by a particular spatial arrangement of the spins at each position $ \bm{x} $.  While there are many possible physical quantum states for a given spin texture, we primarily consider the example of the minimally entangled tensor product of all the individual states $ |\Psi \rangle = \otimes_{\bm{x}} | \psi(\bm{x}) \rangle $.  This realization of a spin texture can be considered without loss of generality as long as the only observables that are measured are local spin expectation values.  An example of such spin texture is those magnetic materials with skyrmions \cite{Muhlbauer915,Nagaosa,Yuki} in helical ferromagnets such as FeCoSi or MnSi, and spinor BECs in anisotropic magnetic fields \cite{PhysRevLett.100.180403}. In such skyrmionic realizations, the product state is a reasonable approximation, since it is mainly characterized by chirality that originates from the lack of inversion symmetry as well as the presence of anisotropic interactions such as Dzyalonshinskii-Moriya interactions \cite{Yuki, fukuda2011quasi,nagaosa2013topological}.  Thus nearest-neighbor exchange interactions can be well approximated by a mean-field approximation.  

\begin{figure}[tb]
\includegraphics[width=\columnwidth]{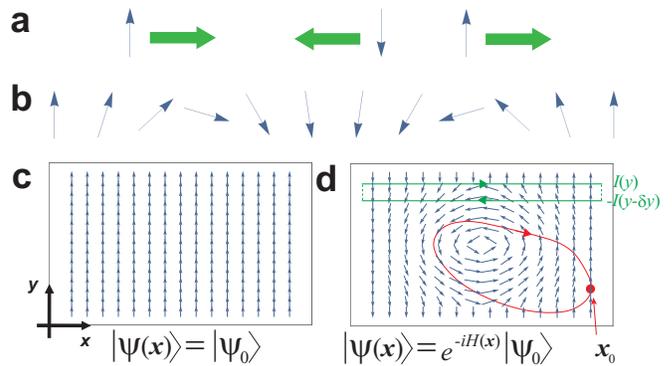}
 \caption{Two types of spin current: (a) spin current due to spatial displacement of spin momenta (large arrows indicate movement); and (b) spin current due to spin waves (spins are fixed in space).  Example of a spin texture in two dimensions: (c) the reference state (``blank canvas'')
$ |\Psi \rangle = \otimes_{\bm{x}} | \psi_0 \rangle $, and (d) the spin texture parametrized by the generator $ G (\bm{x}) $.   }
\label{fig1}
\end{figure}

We parametrize the state of a spin at position $ \bm{x} $ as 
\begin{align}
| \psi(\bm{x}) \rangle = e^{-i G (\bm{x})} | \psi_0 \rangle, 
\label{psidef1}
\end{align}
where $ | \psi_0 \rangle $ is a fixed reference state that the generating operator $ G (\bm{x}) $ transforms into $ | \psi(\bm{x}) \rangle  $.  Thus before the generator acts, the state $ |\Psi \rangle  $ is completely uniform, and is a ``blank canvas'' [see Figs. \ref{fig1}(c) and \ref{fig1}(d)].  Since $ | \psi_0 \rangle $ can be taken arbitrarily, let us take it to be $ | \psi_0 \rangle  = | \psi(\bm{x}_0) \rangle$.  Thus at position $ \bm{x}_0 $, the generator is simply $ G (\bm{x}_0) = 0 $.   We note that, although we refer to $ \bm{x} $ as being ``position,'' in fact it can equally be any other continuous parameter such as momentum depending upon the realization of the spin texture.

From the fundamental theorem of calculus for line integrals,  Eq. (\ref{psidef1}) can be rewritten (see Appendix A for details) as
\begin{align}
| \psi(\bm{x}) \rangle = \exp \left(-i \int_{\cal C}  \bm{\nabla} G (\bm{x}) \cdot d \bm{l} \right) | \psi_0 \rangle, 
\label{psidef}
\end{align}
where $ {\cal C}  $ denotes a contour starting at position $ \bm{x}_0 $ and ending at $ \bm{x} $ (see the solid line in Fig. \ref{fig1}). Taking the line integral to be a loop starting and ending at $ \bm{x}_0 $, we assert that the state taken around the loop $ | \psi(\bm{x}_0) \rangle$ must be exactly equal to the initial state $ | \psi_0 \rangle $. The equality includes global phases, and is an equality in a strict mathematical sense. As the reference state $ | \psi_0 \rangle $ can be taken to be any state, we deduce that for any continuous spin texture we must have
\begin{align}
\exp \left(-i \oint_{\cal C}  \bm{\nabla} G (\bm{x}) \cdot d \bm{l} \right) = I ,
\label{ointeq}
\end{align}
where $ I $ is the $ D $-dimensional identity matrix.  This relation has the meaning that, if we accumulate the changes in the generating operator of a continuous spin texture around a loop, one must necessarily start back at the original state, obviously since it is the same state. 

Let us now illustrate Eq. (\ref{ointeq}) for some simple examples.  The simplest case is where the generator is a U(1) scalar ($ D=1 $) in two dimensions. The solution of Eq. (\ref{ointeq}) is
\begin{align}
\oint_{\cal C}  \bm{\nabla} G (\bm{x}) \cdot d \bm{l}  = 2 \pi m,
\label{vortex}
\end{align}
where $m $ in an integer.  For example, the vortex wavefunction in a BEC, which takes the form $ | \psi(r,\theta) \rangle = e^{i m \theta} | \psi_0 \rangle $, can be easily shown to satisfy Eq. (\ref{vortex}), where $ m $ is the vortex winding number.  For spin $S=1/2$ SU(2) rotations ($ D=2 $), the solution of (\ref{ointeq}) takes the form
\begin{align}
\oint_{\cal C}  \bm{\nabla} G (\bm{x}) \cdot d \bm{l}  = \pi m + \pi m' \bm{u} \cdot \bm{\sigma},
\label{spinvortex}
\end{align}
where $ \bm{u} $ is a unit vector with $ \bm{u}\cdot \bm{u} = 1 $, $ \bm{\sigma} = ( \sigma^x, \sigma^y,\sigma^z) $ are the Pauli matrices, and $ m$ and $m' $ are integers such that $ (-1)^{m+m'} = 1 $. The spin-vortex wave function $ | \psi(\bm{x}) \rangle = e^{i \left(f(\bm{x})+g(\bm{x}) \bm{u} \cdot \bm{\sigma}  \right)} | \uparrow \rangle $ satisfies Eq. (\ref{spinvortex}), with 
the integers $ m$ and $m' $ depending upon the functions $ f(\bm{x})$ and $g(\bm{x}) $. For odd $ m$ and $m' $, half-vortex solutions are obtained \cite{zhou2003quantum}.  

Equation (\ref{ointeq}) may be solved for a $ D $-dimensional system by taking the matrix logarithm of the identity matrix
\begin{align}
\oint_{\cal C} \bm{\nabla} G (\bm{x}) \cdot d \bm{l}  = 2 \pi \sum_{k=1}^D m_k | {\cal C}, k \rangle  \langle {\cal C}, k |  ,
\label{ointsol}
\end{align}
where $ |  {\cal C}, k \rangle $, $ k\in [1,D] $,  are the orthonormal eigenstates of $ \oint_{\cal C}  \bm{\nabla} G (\bm{x}) \cdot d \bm{l} $, and $ m_k $ are integers.  We have thus deduced that all eigenstates must have integer eigenvalues, up to a factor of $ 2 \pi $.

%


\section{\label{sec:quantizedspincurrent}Quantized spin current}
We now would like to relate Eq. (\ref{ointsol}) to some physically observable quantities to obtain the QSHE. Up to this point, no assumptions were made regarding the types of spin textures [and hence the types of generator $ G(\bm{x}) $] that are allowable. To relate the above results regarding the acting generator to the current in a simple way, we assume henceforth that $ [\bm{\nabla} G (\bm{x}),  G (\bm{x}) ] = 0 $.  This condition is satisfied, for example, when the generator is self-commuting at different locations; i.e., it has a fixed axis of rotation.  
Evaluating the current $ \bm{j} (\bm{x}) \equiv -\frac{i \hbar}{2M} ( \langle \psi (\bm{x}) | \bm{\nabla} \psi (\bm{x}) \rangle - \langle \bm{\nabla}  \psi (\bm{x}) | \psi (\bm{x}) \rangle ) $ for Eq. (\ref{psidef1}) yields
\begin{align}
\bm{j} (\bm{x}) = -\frac{\hbar}{M} \langle \psi_0 | \bm{\nabla} G (\bm{x}) | \psi_0 \rangle = -\frac{\hbar}{M} \langle \bm{\nabla} G (\bm{x}) \rangle_0 .
\end{align}
We can then obtain a relation with respect to the current by contour integration:
\begin{align}
\oint_{\cal C} \bm{j} (\bm{x})  \cdot d \bm{l}  = - \frac{h}{M} \sum_k m_k \langle P_k \rangle_0 ,
\label{ointsolproj}
\end{align}
where $ P_k = |  k \rangle  \langle  k | $ is the projection operator onto the $k$th eigenstate. We have dropped the contour label $ {\cal C} $ on the eigenstates, as for the conservative cases the eigenstates are independent on the particular contour. The above does not necessarily lead to a quantized relation for the integrated current around a loop due to the factor of $ \langle P_k \rangle_0 $ which is in general not an integer.

We can however construct a special type of current which does follow a quantization condition.  First define the projected currents, defined as
\begin{align}
\bm{j}_k (\bm{x}) \equiv -\frac{i \hbar}{2M} ( \langle \psi (\bm{x}) | P_k | \bm{\nabla} \psi (\bm{x}) \rangle - \langle \bm{\nabla}  \psi (\bm{x}) |P_k | \psi (\bm{x}) \rangle ),
\end{align}
which follow the relation
\begin{align}
\oint_{\cal C} \bm{j}_k (\bm{x})  \cdot d \bm{l}  = - \frac{h \langle P_k \rangle_0  }{M} m_k .
\label{ointsolquant}
\end{align}
Then using the fact that $ \sum_k P_k = I $, we can deduce that the ``quantized spin current,'' defined as
\begin{align}
\bm{j}_{\text Q} (\bm{x}) =  \bar{m} \sum_k \frac{\bm{j}_k (\bm{x})}{m_k},
\label{quantizedcurrent}
\end{align}
follows
\begin{align}
\oint_{\cal C} \bm{j}_{\text Q} (\bm{x})  \cdot d \bm{l} = - \frac{h}{M} \bar{m},
\label{quantizedoint}
\end{align}
where $ \bar{m} \equiv \prod_k m_k  $ is an integer. The factor of $  \bar{m} $ is placed in Eq. (\ref{quantizedcurrent}) to avoid division by the integer $ m_k $, which are potentially zero. The quantized spin current (\ref{quantizedoint}) depends on the contour $ \cal C $ in a topological way.  The relation is a generalization of the well-known relation that the loop integral of the phase around a vortex in a BEC is an integer multiple of $ 2 \pi $ \cite{pitaevskii03,PhysRevA.92.023629}.  In the same way, Eq. (\ref{quantizedoint}) only depends upon what singularities are circled in the contour, which determines the integers $ m_k $.  For each singularity, there is a set of integers which characterize the nature of the vortex [Fig. \ref{fig1}(d)].  For a contour which does not circle any singularities, $ m_k =0$. 

With this quantized spin current, we can follow an argument similar to that in Ref. \cite{PhysRevA.92.023629} to obtain the total quantized spin current, which is the experimentally observed quantity.  For simplicity we only consider the case that there is only one singularity located at 
$\bm{x} = (x_\text{s},y_\text{s}) $ in the channel, as shown in Fig. \ref{fig1}(d). Now define the quantity
\begin{align}
I(y) = \int_{-\infty}^{\infty} j_{\text Q}^x (\bm{x}) dx,
\end{align}
where $ j_{\text Q}^x  $ is the $x$ component of $ \bm{j}_{\text Q} $. Now consider a long rectangular contour shaped like that shown in Fig. \ref{fig1}(d).  Far away from the vortex, all currents should be zero as there are minimal change spin configurations; thus we can just consider the contributions due to the $ I(y) $.  We thus can deduce that 
\begin{align}
I(y) = \left\{
\begin{array}{ll}
j_0 & \text{ if } y < y_\text{s}, \\
j_0 + \frac{h}{M} \bar{m}  & \text{ if } y > y_\text{s},
\end{array}
\right.
\label{iints}
\end{align}
where $ j_0  = I(0) $ is the current along the bottom edge of the channel.  Using Eq. (\ref{iints}) we can easily evaluate the total quantized spin current
\begin{align}
J_{\text Q}^x & \equiv \int_{-\infty}^\infty j_{\text Q}^x (\bm{x}) dx dy  = 
\int_{-\infty}^\infty I(y) d y  = j_0 w + \frac{h}{M} \bar{m} y_\text{s} ,
\end{align}
where $ w $ is the width of the channel in the $ y$ direction. 
The quantized transverse conductance is obtained by shifting the singularity position and observing the change in the quantized spin current:
\begin{align}
\sigma_{\text Q} = \frac{dJ_{\text Q}^x}{d y_\text{s} }  = \frac{h}{M} \bar{m}  .
\label{qshefinal}
\end{align}
This shows explicitly the QSHE. The remarkable aspect of Eq. (\ref{qshefinal}) is that it relates two experimentally measurable quantities, $ J_{\text Q}^x $ and 
$ y_\text{s} $, to the winding number of the skyrmions.  The current $ J_{\text Q}^x $ is simply a linear combination of spin currents in the $ x $ direction, and $ y_\text{s} $ is the vortex position. Although the quantization condition arises originally from the loop integral (\ref{quantizedoint}), the final expression (\ref{qshefinal}) only involves the quantized current along the $x$ direction, since the rectangular contours of Fig. \ref{fig1}(d) can be evaluated exactly.  Both the current $ J_{\text Q}^x $ and the vortex position $ y_\text{s} $ are readily observable given the images of the spin texture such as that obtained in Ref. \cite{jiang2017direct}.

\begin{figure}[t]
\includegraphics[width=\columnwidth]{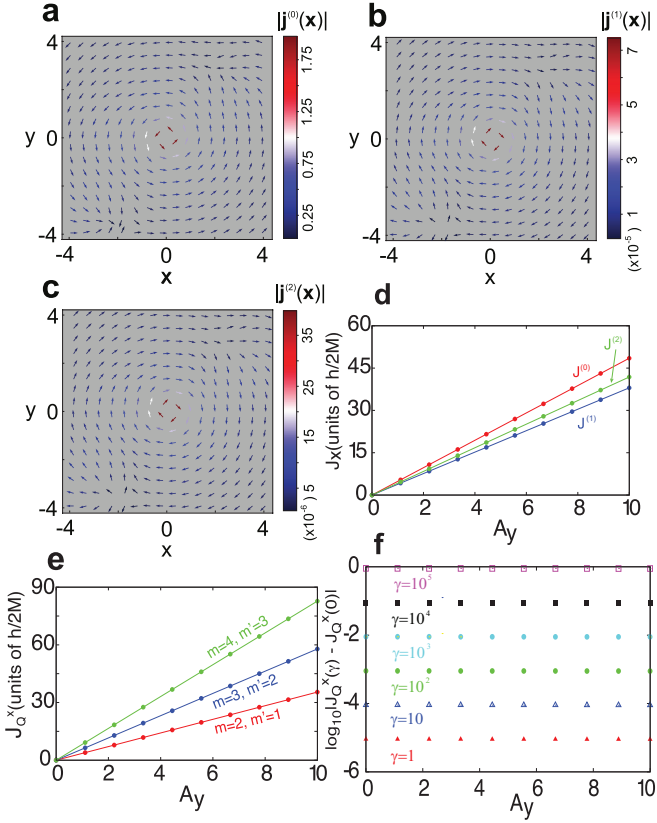}
 \caption{\label{fig2}
Examples of the quantum spin Hall effect using spin textures.  The current distribution in a spin-1 BEC for (a) the conventional current $ \bm{j}^{(0)} ( \bm{x}) $, (b) the first-order spin current $ \bm{j}^{(1)} ( \bm{x}) $, and (c) the second-order spin current $ \bm{j}^{(2)} ( \bm{x}) $.  The chosen spin texture is $f(\bm{x})=m \theta,$ and $g(\bm{x}) = m' \theta n(\bm{x}), $ with $ m = 1 $, $ m' = -2 $, $\bm{u}=(1,1,1)/\sqrt{3}$, and $  n ( \bm{x}) = 1 + \gamma \sum_i \frac{1}{\sqrt{2\pi \sigma ^2}}e^{-\frac{|\bm{x}-\bm{x}_i|^2}{2\sigma ^2}} $ is the noise function with $ \gamma = 0.5 $ and $ \sigma = 1 $.  The noise centers are at $(-2,0)$, $(3,2)$, and $(-3,-3)$.  (d) The $l$th-order integrated spin current $J_x^{(l)} = \int j_x^{(l)} (\bm{x}) dx dy $  versus the asymmetry parameter $A_y$. (e) The total quantized spin current $ J_\text{Q}^x $ versus the asymmetry parameter $ A_y $ for various spin textures for the parameters as marked.  (f) Error in the current due to the inclusion of noise for the case $ m =1$ and $m'=2 $ for the noise strengths as marked.   The Gaussian density profile $ \rho(\bm{x}) = e^{-y^2/(2\sigma_y^2)} $ is assumed with $ \sigma_y = 100  $.  }
\end{figure}

\section{\label{sec:generalizationandexample}Generalization with Spin-$1$ example}
We now illustrate and slightly generalize the theory, using the example of a BEC with spin $S=1 $ spin texture (see Appendix B).  The $S=1 $ case is of particular relevance to BECs with a ground-state hyperfine structure with three hyperfine states, such as $^{87}\text{Rb} $. The spin texture we consider is $  |\psi(\bm{x}) \rangle = e^{i \left(  f(\bm{x})+g(\bm{x}) \bm{u} \cdot \bm{S}  \right) } |\psi_0 \rangle $, where $\bm{S} = (S_x, S_y, S_z) $ are $ S=1 $ spin matrices \cite{Dalibard}. We note that any spin texture of this form satisfies $ [\nabla G(\bm{x}), G(\bm{x}) ] = 0 $. We first construct the quantized spin current, which will be a sum of three currents as given in Eq. (\ref{quantizedcurrent}), with projectors being the eigenstates of $ \bm{u} \cdot \bm{S} $.  An equivalent way of writing the quantized spin current is in terms of conventional and $ l $th-order spin currents which may be evaluated for $ l = 0$, $1$, and $2 $:
\begin{align}
\bm{j}^{(l)} & (\bm{x})  \equiv -\frac{i \hbar}{2M} \langle \psi (\bm{x}) | (\bm{u} \cdot \bm{S})^l | \bm{\nabla} \psi (\bm{x}) \rangle + \text{H.c.} \nonumber \\
&  = \frac{\hbar}{M} \left[ \langle (\bm{u} \cdot \bm{S})^l \rangle_0 \bm{\nabla} f(\bm{x}) + \langle (\bm{u} \cdot \bm{S})^{l+1}  \rangle_0\bm{\nabla} g(\bm{x})  \right] 
\end{align}
with $ \langle (\bm{u} \cdot \bm{S} )^{3} \rangle_0 =  \langle \bm{u} \cdot \bm{S} \rangle_0 $ for $ S = 1 $.  The quantized spin current in terms of these currents is obtained by eliminating the $ \langle (\bm{u} \cdot \bm{S})^l \rangle_0 $ factors:
\begin{align}
\bm{j}_{\text Q} (\bm{x})  = \frac{1}{4} \left[ (m^2 - {m'}^2) \bm{j}^{(0)} (\bm{x})  - m m' 
\bm{j}^{(1)} (\bm{x}) + {m'}^2 \bm{j}^{(2)} (\bm{x}) \right],
\label{quantcurrentbec}
\end{align}
where $ m$ and $m' $ are the winding numbers associated with our chosen functions $ f ( \bm{x}) $ and $ g ( \bm{x}) $ respectively. 
 
Since we are modeling a BEC, we also need to include the density distribution $ \rho( \bm{x} ) $, which has the effect of modulating the current by this factor. In general this acts to spoil the quantization relation (\ref{ointsolquant}) as a factor of $ \rho(x) $ is integrated in the loop integral.  However, using the distribution $ \rho(y) $ in a long channel that is only $ y $ dependent allows for determination of the total current \cite{PhysRevA.92.023629}.  In this situation the total quantized spin current is
\begin{align} 
J^x_{\text{Q}} &=\frac{\hbar P_x \mathcal{N}}{M} +\frac{h}{2M} \left(m^2-{m'}^2 \right) A_y,
\label{totalcurrentbec}
\end{align}
where $\mathcal{N}$ is the total number of particles in the BEC,  $P_x$ is related to the center of mass momenta of the condensate, and $A_y$ is the spatial asymmetry parameter $ A_y =\int_{\infty}^{y_\text{s}}\rho(y) dy -\int_{y_\text{s}}^{\infty}\rho(y) dy $. 
Equation (\ref{totalcurrentbec}) is our desired result, where we explicitly have a quantized $ d J^x_{\text{Q}} / d A_y $.

We numerically simulate the spin texture $ |\psi(\bm{x}) \rangle $ for the functions $ f ( \bm{x}) = m \theta $ and $ g ( \bm{x}) = m' \theta  n ( \bm{x}) $, respectively. Here, $  n ( \bm{x}) $ is a noise function which acts to locally disrupt the vortex distribution but does not affect the overall topology of the state [see Figs. \ref{fig2}(a)-\ref{fig2}(c)].  
For each spin configuration, we calculate the total quantized spin current and asymmetry parameter by integrating Eq. (\ref{quantcurrentbec}) and $ A_y $, respectively. Different values of $ J^x_{\text{Q}}$ and $A_y $ are obtained by displacing the density distribution $ \rho( \bm{x}) $ in the $ y $ direction and keeping  $ |\psi(\bm{x}) \rangle $ fixed. 

Calculated results for different winding numbers $m$ and $m'$ are shown in Figs. \ref{fig2}(d) and \ref{fig2}(e), which represent the analogous results as found in Ref.  \cite{PhysRevA.92.023629} where the displacement of a vortex in the $ y $ direction gives rise to a current in the $ x $ direction. Here we also see a perfectly linear relationship between $ J^x_{\text{Q}} $ and $A_y$ as expected, but the current is both a combination of the spin and conventional currents, which occurs due to each of them contributing their own quantization relation.  Only the quantized spin current obeys a quantized QSHE as this is a topological invariant of the system.  As expected, the results are very robust in the presence of noise.  In Fig. \ref{fig2}(f) we show the logarithmic difference between the currents in the presence of noise of various strengths.  The results show that one must add extremely large amounts of noise before the current-asymmetry relation is disrupted.   The robustness can be attributed to the fact that the observables are connected to a topological invariant, which is insensitive to local fluctuations.  

\section{\label{sec:summary}Summary and conclusions}
In summary, we have shown how to obtain a skyrmionic QSHE in spin textures, in contrast to the conventional approach of employing spin-orbit interactions.  The result is a general property resulting from the single-valuedness of the wavefunction at each point in space -- tracking the changes of the state around a loop must result in the same state again.  In this case a combination of spin and conventional currents must be used to obtain a precise quantized conductance.  This does require {\it a priori} knowledge of the topological state of the system -- however, since the combinations only involve coefficients that are integers, these should be estimatable fairly easily if images of the spin texture are available.  The primary assumption made in our derivation is that $ [\nabla G ( \bm{x} ), G ( \bm{x} ) ] = 0 $, which is satisfied by ``single-axis'' generating operators.  Since this still allows for a completely general spatial distribution, this still gives a wide variety of spin textures, including the presence of noise.  We found numerically that there is very little influence of noise on the conductance, as would be expected from a QHE. Experimentally, the method requires the ability to image and manipulate the skyrmion position, which has been shown to be possible in a variety of systems \cite{jiang2017direct,litzius2017skyrmion,samson2016deterministic}.

\begin{acknowledgments}
This work is supported by the Shanghai Research Challenge Fund; New York University Global Seed Grants for Collaborative Research; the National Natural Science Foundation of China (Grant No.$61571301$ and No.$\text{D}1210036\text{A}$); the NSFC Research Fund for International Young Scientists (Grant No.$11650110425$ and No.$11850410426$); NYU-ECNU Institute of Physics at NYU Shanghai; the Science and Technology Commission of Shanghai Municipality (Grant No.$17\text{ZR}1443600$); the China Science and Technology Exchange Center (Grant No.$\text{NGA}-16-001$); and an NSFC-RFBR Collaborative grant (Grant No.$81811530112$).
\end{acknowledgments}

\appendix
\section{\label{app: eq2}Line integral form of spin state}
We here include the derivation of Eq. \eqref{psidef}. We start with the parametrization of the spin texture
\begin{align}
| \psi(\bm{x}) \rangle = e^{-i G (\bm{x})} | \psi_0 \rangle. 
\label{spintext}
\end{align}
Now expand the generating operator in terms of an operator basis $ A_j $:
\begin{align}
G (\bm{x}) = \sum_j a_j (\bm{x}) A_j ,
\label{gexp}
\end{align}
where $ a_j (\bm{x}) $ are position-dependent scalar functions. Applying the fundamental theorem of 
calculus for line integrals for the scalar functions $ a_j (\bm{x}) $ we have
\begin{align}
 a_j (\bm{x}) -  a_j (\bm{x}_0) = \int_{\cal C}  \bm{\nabla}  a_j (\bm{x}) \cdot d \bm{l} ,
\label{fundcalc}
\end{align}
where the contour starts at $\bm{x}_0 $.  Since $ G(\bm{x}_0) = 0 $, we may take $ a_j (\bm{x}_0) = 0 $.  Substituting Eqs. (\ref{fundcalc}) and (\ref{gexp}) into Eq. (\ref{spintext}), we obtain
\begin{align}
| \psi(\bm{x}) \rangle = \exp \left(-i \int_{\cal C}  \bm{\nabla} G (\bm{x}) \cdot d \bm{l} \right) | \psi_0 \rangle .
\end{align}

\section{\label{app:spinhalfexample} spin-$1/2$ example}

In the spin-1/2 case, we can show explicitly that the generating operator from Eq. (\ref{psidef1}) in the main text is
\begin{align} \label{psigen}
| \psi (\bm{x}) \rangle = e^{i \left(f(x,y)+g(x,y) \left (\mathbf{u}\cdot \mathbf{\sigma}  \right) \right)} |\psi_0 \rangle 
\end{align}
where $f(x,y)$ and $g(x,y)$ are scalar functions, and $\bm{u}$ is an arbitrary unit vector. The current density $ \bm{j} (\bm{x}) = -\frac{i \hbar}{2M} ( \langle \psi (\bm{x}) | \bm{\nabla} \psi (\bm{x}) \rangle - \langle \bm{\nabla}  \psi (\bm{x}) | \psi (\bm{x}) \rangle ) $ can be evaluated for Eq. (\ref{psigen}) to give 
\begin{align} \label{eq:j_x}
\bm{j} (\bm{x}) = \frac{\rho \hbar}{M}\left[ \bm{\nabla} f(x,y) +\bm{\nabla} g(x,y) \langle \bm{u} \cdot \bm{\sigma} \rangle_0 \right ] ,
\end{align}
where the expectation value is with reference to $ | \psi_0 \rangle $.
A similar expression for spin current density $\bm{j} (\bm{x}) = -\frac{i \hbar}{2M} ( \langle \psi (\bm{x}) |(\bm{u} \cdot \bm{\sigma}) |\bm{\nabla} \psi (\bm{x}) \rangle - \langle \bm{\nabla}  \psi (\bm{x}) |(\bm{u} \cdot \bm{\sigma})  |\psi (\bm{x}) \rangle )$ can be written as:
\begin{align}\label{eq:j_usigma}
\bm{j}_{(\bm{u}\cdot \bm{\sigma})} (\bm{x}) =\frac{\rho \hbar}{M}\bigg[  \bm{\nabla} f(x,y) \langle \bm{u} \cdot \bm{\sigma} \rangle_0 +\bm{\nabla} g(x,y) \bigg] 
\end{align}

We can then obtain a total current which is denoted as $\bm{J}_{\text{tot}}$ by integrating both sides of Eqs. \eqref{eq:j_x} and \eqref{eq:j_usigma}:
\begin{align}
\bm{J}_{\text{tot}} &=N_f \oint_{\cal C} \bm{j} (\bm{x})\cdot d\bm{l}-N_g \oint_{\cal C} \bm{j}_{(\bm{u}\cdot \bm{\sigma})} (\bm{x}) \cdot d\bm{l} \nonumber \\ &=\frac{\rho h}{M}\left (N_f^2-N_g^2 \right).
\end{align}
We choose $f(x,y)=N_f\bm{\theta}$ and $g(x,y)=N_g\bm{\theta}$ so that they satisfy the phase-winding relation of BEC: $\oint_{\cal C}\nabla f(x,y)\cdot d\bm{l}=2\pi N_f$, $\oint_{\cal C}\nabla g(x,y)\cdot d\bm{l}=2\pi N_g$. The property of Pauli matrices $\langle(\bm{u} \cdot \bm{\sigma})^2 \rangle_0 =1$ is used. These results can be taken as a generalization of current density in the previous work \cite{PhysRevA.92.023629}. 



\end{document}